\documentclass[conference]{IEEEtran}
\usepackage[utf8]{inputenc}
\usepackage{float}
\usepackage{graphicx}
\usepackage{epstopdf}
\usepackage{colortbl}
\usepackage{listings}
\usepackage{color}
\usepackage{multirow}
\usepackage{epstopdf}
\usepackage{algorithm}
\usepackage{algorithmic}
\usepackage{tcolorbox}
\usepackage{multirow}
\usepackage{mathtools}
\usepackage{amsfonts}
\usepackage{xspace}
\usepackage[hyphens]{url}
\usepackage{dblfloatfix}
\usepackage{array}
\usepackage{enumerate}
\usepackage{subcaption}
\usepackage[font=small,labelfont=bf]{caption}
\usepackage[normalem]{ulem}

\usepackage{pgfplots}
\usetikzlibrary{matrix}
\usepgfplotslibrary{groupplots}
\pgfplotsset{compat=newest}
\pgfplotsset{compat=newest,
legend style={font=\footnotesize},
label style={font=\footnotesize},
tick label style={font=\footnotesize},
title style={font=\footnotesize}}
\usepackage{pifont}
\usepackage{marvosym}
\usepackage{tikz}
\usetikzlibrary{shapes,arrows}
\usetikzlibrary{backgrounds, positioning, fit}

\begin{document}

\title{Stealing Neural Networks via Timing Side Channels}

\author{
    \IEEEauthorblockN{Vasisht Duddu\IEEEauthorrefmark{1}, Debasis Samanta\IEEEauthorrefmark{2}, D. Vijay Rao\IEEEauthorrefmark{3},  Valentina E. Balas\IEEEauthorrefmark{4}}
    \IEEEauthorblockA{\IEEEauthorrefmark{1}Indraprastha Institute of Information Technology, Delhi, India}
    \IEEEauthorblockA{\IEEEauthorrefmark{2}Indian Institute of Technology, Kharagpur, India}
    \IEEEauthorblockA{\IEEEauthorrefmark{3}Institute for Systems Studies and Analyses, Delhi, India}
    \IEEEauthorblockA{\IEEEauthorrefmark{4}Aurel Vlaicu University of Arad, Arad, Romania}
    \IEEEauthorblockA{vduddu@tutamail.com, dsamanta@iitkgp.ac.in, doctor.rao.cs@gmail.com, valentina.balas@uav.ro}
}


\maketitle

\begin{abstract}
Deep learning is gaining importance in many applications. However, Neural Networks face several security and privacy threats. This is particularly significant in the scenarios where Cloud infrastructures deploy a service with Neural Network models at the back end. Here, an adversary can extract the Neural Network parameters, infer the regularization hyperparameter, identify if a data point was part of the training data, and generate effective transferable adversarial examples to evade classifiers. This paper shows how a Neural Network model is susceptible to timing side channel attack. In this paper, a black box Neural Network extraction attack is proposed by exploiting the timing side channels to infer the depth of the network. Although, constructing an equivalent architecture is a complex search problem, it is shown how Reinforcement Learning based optimisation can efficiently reduce the search space and reconstruct optimal substitute architecture close to the target model. The proposed approach has been tested with VGG architectures on CIFAR10 data set. It is observed that it is possible to reconstruct substitute models with test accuracy close to the target models and the proposed approach is scalable and independent of type of Neural Network architectures.
\end{abstract}


\section{Introduction} \label{sec:intro}

Of late, Neural Networks have been successfully employed to many diversified areas, namely computer vision, natural language processing and business intelligence \cite{bishop2006pattern}.
Deep learning architectures have also been deployed for automating critical decision making in security applications like national critical infrastructures, malware and intrusion detection \cite{Xu2016AutomaticallyEC}.
For various military applications such as unmanned combat aerial vehicle, automated target recognition and guided missile systems, the underlying decision making depends on state of the art deep learning architectures.
Banks and financial services rely on deep learning techniques to process massive financial data. Autonomous driving has attracted several big automotive companies like Audi, Tesla and Waymo to invest billions of dollars into Deep Learning Research.

Designing and engineering Neural Networks for commercial services requires significant time, money and human effort, ranging from collection of massive data to fine-tuning the hyperparameters of the model for performance improvement.
The commercial value of these models make them an important intellectual property for companies, due to which the model attributes such as number of layers, training algorithms and regularisation hyperparameter, are kept confidential as a black box.
These black box models do not reveal any information to the service users other than the output predictions for the corresponding input.
This has been commercialised as a business model by several cloud service providers such as Google, Amazon, Microsoft and BigML by deploying an end-to-end infrastructure for using Deep Neural Networks as a service.
In Machine Learning as a Service (MLaaS) paradigm, trusted users submit training data to the service providers, who spend significant resources to design and train high performing models, deployed for public use on a pay-per-query basis.
Despite its promise, the commercial value of the black box Neural Networks within MLaaS makes them susceptible to adversary's attempts to extract the model functionality, model attributes, and circumventing the pay-per-query setting of the service.
Given the architecture of Neural Networks, an adversary can further mount various privacy and security attacks like model inversion \cite{fredrikson2015model}\cite{hitaj2017deep} and membership inference \cite{shokri2017membership}\cite{salem2018ml} to infer the input and training data instances and generate more accurate adversarial examples to evade classifiers during test phase \cite{Papernot:2017:PBA:3052973.3053009}.
These attacks violate the privacy of the sensitive data used for training the models and provide a way for the adversaries to evade security systems such as malware and intrusion detection systems thereby forcing to make incorrect predictions.

A major security question in such a black box setting such as MLaaS addressed in this paper is \textit{Can a weak adversary in a black box setting efficiently infer target Neural Network attributes by exploiting side channels with minimum number of queries?}
In this work, a novel model extraction attack in a black box setting is proposed by exploiting timing side channels and efficiently reconstructing a substitute model architecture with functionality close to the target model using a constant number of queries.

\textbf{Key Challenges in Model Extraction Attacks.} Stealing a Neural Network architecture and its functionality is a challenging problem owing to the large number of hyperparameters making brute-force infeasible.
The rapid growth of Neural Network design space has increased the complexity of architectures making the problem of black-box model extraction more challenging.
In the black box setting as in MLaaS, the adversary has only access to the output predictions given input and lacks any knowledge about the model and the training data.
Previously, model extraction attacks have relied on using input-output relations to identify the decision boundary of the target model \cite{tramer2016stealing}\cite{oh2018towards}.
However, such attacks require significant computational resources and incur a huge time overhead to search for the substitute model.
For instance, given a prior knowledge about the number of layers and type of layers in a Neural Network, it still takes 40 GPU days to search for a simple 7 layer networks architecture \cite{oh2018towards}.
Further, these attacks do not accommodate the state of the art architectures with complex topologies and skip connections \cite{DBLP:journals/corr/HuangLW16a}.
While extracting the model, these attacks require large number of queries which grow with the size of the architecture making the attacks highly inefficient \cite{milli2018model}\cite{tramer2016stealing}.
An alternative approach for model extraction is to exploit side channels like power consumption \cite{cryptoeprint:2018:477}, memory access patterns \cite{Hua:2018:REC:3195970.3196105}\cite{DBLP:journals/corr/abs-1903-03916} and cache side channel attacks \cite{yan2018cache}\cite{hong2018security} to infer target model attributes.
While these attacks give fine grained information about the target model during execution, the threat model requires escalated adversary privileges and strong assumptions like physical access to hardware and shared resources for processes running on the server.

\textbf{Proposed Approach.} The objective of model extraction attack is to search for a substitute model with similar functionality as the target neural architecture. However, the search space for the substitute model is very large and complex due to the large number of hyperparameters in Neural Networks.
To make the search tractable and efficient, the adversary has to reduce the search space by identifying some of the attributes of the target Neural Network in a black box setting using minimum queries. In a black box setting, a weak adversary can obtain the output prediction corresponding to a given input image.
This paper shows the existence of timing side channels in the black box setting due to the dependence of the total execution time of the Neural Network on the total number of layers or depth of the network.
From the total execution time, an adversary can infer the total number of layers (depth) of the Neural Network using a regressor trained on the data containing the variation of execution time with Neural Network depth.
This additional side channel information obtained, namely the depth of the network, reduces the search space for finding the substitute model with functionality close to the target model.

To efficiently search for the optimal Neural Network, an optimisation problem is introduced which is solved using Reinforcement Learning based Neural Architecture Search.
A Recurrent Neural Network(RNN) controller predicts a new substitute architecture, whose performance determines the reward to improve the controller's prediction for subsequent optimisation epochs \cite{DBLP:journals/corr/ZophL16}.
The optimisation problem involves minimising the loss function computed between the predicted labels of the substitute model and the target model instead of the true labels \cite{hinton2015distilling}.
This ensures that the substitute model learns to mimic the predictions of the target model and hence, increasing the similarity of the two models.
Over multiple epochs of training, the RNN controller updates its parameters based on the reward obtained by minimising the loss, to predict optimal substitute architectures with performance close to target model.
The proposed architecture search and reconstruction technique can be used with other attack approaches as well, like cache side channel attacks \cite{hong2018security}\cite{yan2018cache}.

The proposed approach assumes a weak adversary with only black box access to the target Neural Network and requires a constant number of queries to infer the Neural Network depth independently of the architecture size.
Further, the objective function of Reinforcement Learning maximises the test accuracy of the proposed Neural Network which ensures that the final substitute Neural Network is optimal and close to the target model.

\pgfdeclarelayer{background}
\pgfdeclarelayer{foreground}
\pgfsetlayers{background,main,foreground}

\tikzstyle{image} = [rectangle, rounded corners, minimum width=1.5cm, minimum height=1.5cm,text centered, draw=black, fill=gray!15]
\tikzstyle{time} = [rectangle, minimum width=9.5cm, minimum height=0.5cm,text centered, draw=black]
\tikzstyle{input} = [rectangle, rounded corners, minimum width=0.15cm, minimum height=1.5cm,text centered, draw=black, fill=gray!15]
\tikzstyle{predictions} = [rectangle, rounded corners, minimum width=0.5cm, minimum height=2cm,text centered, draw=black, fill=gray!15]
\tikzstyle{layer1} = [rectangle, rounded corners, minimum width=0.25cm, minimum height=1.5cm,text centered, draw=black, fill=gray!15]
\tikzstyle{layer2} = [rectangle, rounded corners, minimum width=0.25cm, minimum height=1cm,text centered, draw=black, fill=gray!15]
\tikzstyle{layer3} = [rectangle, rounded corners, minimum width=0.25cm, minimum height=0.6cm,text centered, draw=black, fill=gray!15]
\tikzstyle{layer4} = [rectangle, rounded corners, minimum width=0.25cm, minimum height=0.25cm,text centered, draw=black, fill=gray!15]
\tikzstyle{write} = [rectangle, rounded corners, minimum width=2.5cm, minimum height=1.5cm,text centered, draw=black, fill=gray!15]
\tikzstyle{discriminator} = [rectangle, rounded corners, minimum width=3cm, minimum height=1.2cm,text centered, draw=black, fill=gray!15]
\tikzstyle{regression} = [rectangle, rounded corners, minimum width=3cm, minimum height=1.2cm,text centered, draw=black, fill=gray!15]

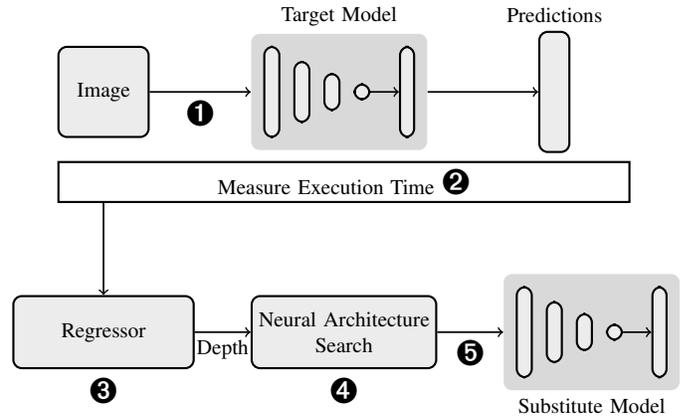
\begin{figure}[t!]
\centering
\resizebox{.5\textwidth}{!}{
\begin{tikzpicture}[node distance=2cm, line width=1pt,every node/.style={align=center}]

\node (en1) [layer1]  at (-8.2,1) {};
\node (en2) [layer2]  at (-7.7,1) {};
\node (en3) [layer3]  at (-7.2,1) {};
\node (en4) [layer4]  at (-6.7,1) {};
\node (pred) [input]  at (-5.95,1) {};

\begin{scope}[on background layer]
    \node (target) [fit=(en1) (en2) (en3) (en4) (pred), fill= gray!30, rounded corners, inner sep=.2cm, label={above:Target Model}] {};
\end{scope}

\node (outpred) [predictions, label={above: Predictions}] at (-3.5,1) {};
\node (img) [image] at (-11,1) {Image};
\node (time) [time] at (-7,-0.5) {Measure Execution Time \LARGE \ding{203}};

\node (en11) [layer1]  at (-4,-3) {};
\node (en12) [layer2]  at (-3.5,-3) {};
\node (en13) [layer3]  at (-3,-3) {};
\node (en14) [layer4]  at (-2.5,-3) {};
\node (pred1) [input]  at (-1.75,-3) {};

\begin{scope}[on background layer]
    \node (substitute) [fit=(en11) (en12) (en13) (en14) (pred1), fill= gray!30, rounded corners, inner sep=.2cm, label={below:Substitute Model}] {};
\end{scope}

\node (controller) [discriminator,label=below:\LARGE \ding{205}] at (-7,-3) {Neural Architecture\\ Search};
\node (reg) [regression, label=below:\LARGE \ding{204}] at (-11,-3) {Regressor};

\draw[thick,->] (en4.east) -- node[anchor=north] {} (pred.west);
\draw[thick,->] (en14.east) -- node[anchor=north] {} (pred1.west);

\draw[thick,->] (img.east) -- node[anchor=north] {\LARGE \ding{202}} (target.west);
\draw[thick,->] (target.east) -- node[anchor=north] {} (outpred.west);
\draw[thick,->] ([xshift=-4cm]time.south) -- node[anchor=north] {} (reg.north);

\draw[thick,->] (reg.east) -- node[anchor=north] {Depth} (controller);
\draw[thick,->] (controller.east) -- node[anchor=north] {\LARGE \ding{206}} (substitute);

\end{tikzpicture}
}
\caption{\textbf{Model Extraction using Timing Side Channels.} 1) The adversary queries the target model by sending an input; 2) Adversary measures the execution time of the target model for the provided input; 3) The execution time is passed to the adversary's regressor model which predicts the depth of the target model; 4) The depth of the Neural Network is used to reduce the search space for the architecture search; 5) Architecture Search produces the optimal architecture using Reinforcement Learning within the constrained search space which has very close test accuracy as the target model.}
\label{overview}
\end{figure}

\textbf{Evaluation.} To measure the success of the proposed model extraction attack, the performance of the regressor to correctly infer the depth of the Neural Network given the total execution time is shown.
The performance of different regressors, to infer the Neural Network depth has been assessed based on the $R^2$ score and the Mean Squared Error to select the regressor model which captures the maximum variance and accuracy.
From the results, ensemble based regressors like random forrest and boosted decision trees outperform their linear counterparts: Ridge regression, Support Vector Machine (SVM) and decision trees.
This is followed by the evaluation of the Reinforcement Learning based architecture search by comparing the test accuracy of the reconstructed model with the target model accuracy.
The experiments are performed using deep convolutional Neural Networks similar to VGG architectures \cite{DBLP:journals/corr/SimonyanZ14a}.
The proposed Reinforcement Learning based architecture search technique can generate a model with test accuracy within 5\% of the target model.

\textbf{Main Contributions.} The paper makes the following main contributions:
\begin{itemize}
\item Shows that Neural Networks are vulnerable to timing side channel attacks as Neural Network architectures with different depth have different execution time (Section~\ref{sec:analysis}).
\item Proposes a novel attack to infer the depth of the Neural Network using timing side channels in constant number of queries in a black box setting (Section~\ref{sec:attack}).
\item Proposes an efficient search technique to reconstruct an optimal substitute architecture using Reinforcement Learning while ensuring functionality similar to the target model (Section~\ref{sec:evaluation}).
\end{itemize}

\section{Background}

\subsection{Neural Networks}

Let $(x,y)$ be the data points obtained by sampling from a probability distribution $D$ over the space $X$ of
input feature values and space $Y$ of output labels. The goal of machine learning algorithms is to learn the mapping from $X$ to $Y$ captured by a function $f:X \rightarrow Y$.
The associated loss function $l_f:X \times Y \rightarrow \mathbb{R}$ captures the error made by the prediction $f(x)$ when the true label is $y$.
Deep Learning, a subset of machine learning algorithms, uses Deep Neural networks modelled as function $f_h(x,\theta)$ where $\theta$ are the parameters optimised during training to obtain minimum expected loss under the constraint of the hyperparameters $h$.
The set of hyperparameters $h$ includes the depth of the Neural Network, stride and filter size of convolution and maxpool layers and regularization hyperparameters.
The performance of the Neural Networks is measured by computing the accuracy on test data in classification tasks.

\subsection{Security and Privacy in Machine Learning}

Machine learning is known to have several security and privacy issues in adversarial settings \cite{DSJ12371}.
A major security threat in Neural Networks is Adversarial examples, i.e, perturbed data instances that fool the classifier into misclassifying the image by either poisoning the training data or evading the decision logic during inference \cite{Papernot:2017:PBA:3052973.3053009}.
For a Neural Network function $f_{h}(x;\theta)$ with input data point $x$ and parameters $\theta$, an adversary can violate the confidentiality and privacy of input data ($x$), training data, model parameters ($\theta$) and the model computation ($f_{h}(x;\theta)$).
The privacy of input passed to the model can be violated using model inversion attack \cite{fredrikson2015model} by extracting the input when the adversary knows the output, parameters and gradients.
Another major class of privacy attacks is Membership inference which violate the privacy of individual members of the training dataset by identifying whether a given data point is in the dataset or not \cite{shokri2017membership}\cite{salem2018ml}.
Further, the computation of the Neural Networks leak information in the form of side channels which allow the adversary to extract model details or inputs \cite{cryptoeprint:2018:477}\cite{wei2018know}.
Machine learning models can be extracted by adversary to reconstruct a substitute model with similar functionality as the target model and hence violating the intellectual property of the service provider \cite{tramer2016stealing}\cite{wei2018know}\cite{wang2018stealing}.
However, some of these attacks assume that the underlying Neural Network architecture is known to the adversary.
Hence, extracting the target model architecture enables the adversary to mount further security and privacy attacks.

Implementation and physical characteristics of systems expose information about the underlying computation which can be extracted in the form of side channels.
Power consumption and Timing side channels are some common manifestations of side channel attacks overlooked during system design. While side channels like power channels
are accurate and reveal significant information about the target architecture, they require expensive equipment and probes to monitor and measure the power \cite{Kocher:1999:DPA:646764.703989}.
Timing Channels arise when the program uses conditional statements dependent on the secret information which influences the runtime or when the access timing correlates strongly with the program locality dependent on the secret \cite{Kocher:1996:TAI:646761.706156}.

\section{Problem Statement}

\textbf{Model Extraction.} Given a black box access to a target Neural Network $f_{target}$, the goal of the adversary is to search for a substitute Neural Network $f_{substitute} \in S$, where $S$ is the search space for all possible Neural Network models with different hyperparameters, such that the functionality of $f_{target}$ approximates $f_{substitute}$ using minimum possible queries. The metric used to measure the functionality of the models, $f_{substitute}$ and $f_{target}$, is the test accuracy ($R_{test}$). In other words, the objective is to minimize the difference in test accuracy $R_{test}=\frac{1}{\vert D \vert} \sum_{(x,y) \in D} d(f_{target}(x),f_{substitute}(x))$ between the two models $f_{target}$ and $f_{substitute}$ for inputs ($x,y$) sampled from the data ($D$), i.e, $(x,y)\stackrel{i.i.d}{\sim}D$ and a given distance function $d$ between the two inputs.

\textbf{Exploiting Timing Side Channels.} To reduce the entropy of search space of possible Neural Network models and make the search more efficient, the adversary exploits the timing side channels to infer the number of layers from the total execution time. For this, the adversary collects a dataset ($D_A$) with execution time ($T$) and Neural Network depth ($K$) for various models by varying the number of parameters. Formally, given the attacker dataset $D_A$ = \{$(T_1 ,K_1 ), \cdots ,(T_N ,K_N )\}$ of i.i.d. random variables, for a given depth of the target Neural Network ($k$) from the total execution time ($t$), the adversary estimates the regression function $R(k) = E\{K \vert T = t\}$.

\textbf{Model Search.} The estimated depth is used to constraint the search space to $S_k$ which is the set of all the Neural Network models of depth $k$. This allows the adversary to search for the substitute model $f_{substitute}$ in the search space $S_k$ instead of search space $S$ where $S_k \subset S$. However, the search space $S_k$ is parameterised by kernel size, stride and number of filters which still make the search space large. To make the search space tractable, the adversary uses Reinforcement Learning paradigm where the accuracy of the target model is included in the objective function as part of the reward, to search for Neural Networks with higher test accuracy.

\section{Threat Model}\label{setting}

\pgfplotsset{footnotesize,height=5cm,width=0.33\textwidth}

\begin{figure*}[!b]
\begin{center}
\resizebox{2\columnwidth}{!}{%
\begin{tabular}{lllll}

\begin{tikzpicture}
\begin{axis}[title={(a)},
title style={at={(0.5,0)},anchor=north,yshift=-20},
legend style={font=\tiny},
legend pos =  north west,
legend entries={Kernel Size=3,Kernel Size=5},
ylabel={Time(ms)},
xlabel={Number of Filters},
grid=major
]
\addplot[
    color=black,
    dotted,
    mark=*,
    mark options={solid},
    smooth
    ]
    coordinates {
	(16,0.747)(32,0.988)(64,1.149)(128,2.2237)(256,4.3379)
    };
\addplot[
      color=black,
      dashed,
      mark=*,
      mark options={solid},
      smooth
    ]
    coordinates {
	(16,0.977)(32,1.561)(64,2.404)(128,3.368)(256,6.244)
    };
\end{axis}
\end{tikzpicture} &

\begin{tikzpicture}
\begin{axis}[title={(b)},
title style={at={(0.5,0)},anchor=north,yshift=-20},
legend style={font=\tiny},
legend pos =  north west,
legend entries={Kernel Size=3,Kernel Size=5},
ylabel={Time(ms)},
xlabel={Number of Filters},
grid=major
]
\addplot[
    color=black,
    dotted,
    mark=*,
    mark options={solid},
    smooth
    ]
    coordinates {
	(16,0.254)(32,0.463)(64,0.9529)(128,1.883)(256,4.0349)
    };
\addplot[
      color=black,
      dashed,
      mark=*,
      mark options={solid},
      smooth
    ]
    coordinates {
	(16,0.5359)(32,1.0569)(64,1.921)(128,3.9649)(256,7.2109)
    };

\end{axis}
\end{tikzpicture} &

\begin{tikzpicture}
\begin{axis}[ylabel={Time(ms)},
title style={at={(0.5,0)},anchor=north,yshift=-20},
title={(c)},
xlabel={Number of Filters},
legend style={font=\tiny},
legend pos =  north west,
legend entries={Stride=1,Stride=2,Stride=3},
grid=major]
\addplot [
    color=black,
    solid,
    mark=*,
    mark options={solid},
    smooth
]
coordinates{
 (16,0.5359)(32,1.056)(64,1.921)(128,3.964)(256,7.210)
};

\addplot [
    color=black,
    dotted,
    mark=*,
    mark options={solid},
    smooth
]
coordinates{
 (16,0.145)(32,0.2749)(64,0.534)(128,0.980)(256,1.7819)
};

\addplot [
    color=black,
    dash pattern=on 1pt off 3pt on 3pt off 3pt,
    mark=*,
    mark options={solid},
    smooth
]
coordinates{
 (16,0.098)(32,0.144)(64,0.27399)(128,0.4829)(256,0.9389)
};
\end{axis}
\end{tikzpicture}
\end{tabular}
}
\caption{(a) Convolution Layer: Total Execution Time increases linearly with kernel size and filter size; (b) Maxpool Layer: Total Exeuction Time increases linearly with kernel and filter size; (c) Maxpool Layer: Total Execution Time decreases with increase in stride.}
\label{fig:arch_plain}
\end{center}
\end{figure*}
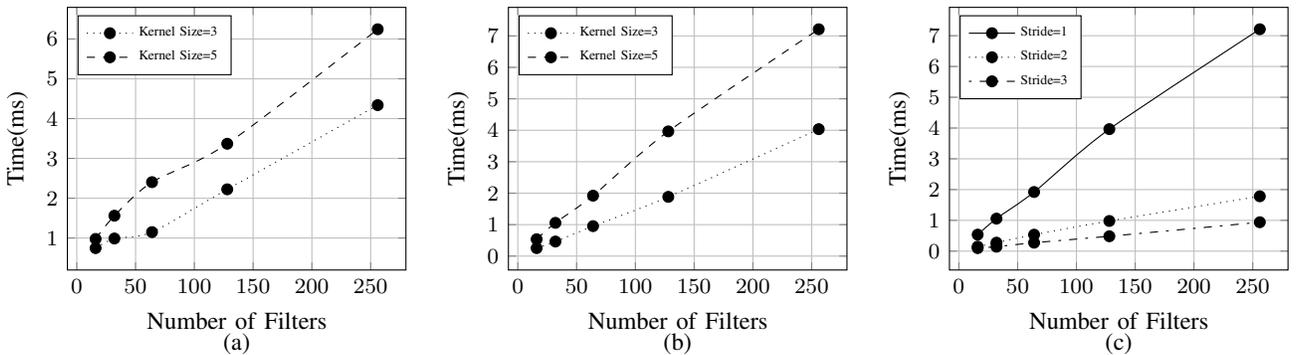

\textbf{Setting.} There are two settings for machine learning in adversarial setting: white box and black box, based on the adversary's knowledge about the target system. In white box setting, the adversary has access to the underlying data, learning algorithms, architecture of the model, training parameters and target model architectures which allows the adversary to compute the output of the intermediate layers. The proposed attack is in a black box setting where the adversary is weak and does not have access to the model internals and can only query the trained model and obtain the corresponding output predictions.

\textbf{Hardware.} The proposed attack is an inference phase attack, where the trained target model has been deployed as a service to the users. During inference, CPUs and Neural Network accelerators are extensively used while GPUs are used for training Neural Network architectures \cite{DBLP:journals/corr/SzeCYE17}. Majority of machine learning Cloud service providers use CPUs \cite{hazelwood2018applied}. The attack is evaluated using CPUs but the approach can be extended to other hardware accelerators as well. The adversary requires the same processor as the target model which can be openly obtained in most of the ML as a Service (MLaaS) platforms like Amazon Sagemaker and Facebook which heavily rely on CPUs for the inference and provide the hardware specifications \cite{amazonec2}\cite{amazonsage}.
The target hardware or the service can be purchased by adversary to run the queries and generate the attack dataset which is a one time operation and done as part of the setup phase for the attack.

\textbf{Data.} The attack assumes a weak adversary with no knowledge about the training data and only knows the input-output dimensions and range of values they can take. The attacker, however, is assumed to know the underlying data distribution from which the training data was sampled. This allows the attacker to sample data points as inputs and pass them to the target model for predictions. There are two main approaches to reconstruct the training data for the substitute model: Iterative membership inference attacks and data reconstruction attacks. In case of membership inference attack, the adversary samples data points from the underlying data distribution $x \sim D$ which is passed as a query to the target model from which the adversary obtains the model output posterior $f(x)$. Given the output posterior of the model for the input, the attacker checks if the value of the maximum posterior is greater than a threshold($x$ $\in$ $D_T$) or not($x$ $\not\in$ $D_T$) \cite{salem2018ml}. This is done iteratively by sampling data points and using membership inference attacks to reconstruct the training data used by the target Neural Network. In data reconstruction attack, an adversary uses a generative adversarial network to reconstruct training data samples from target model by finding the approximate training data distribution \cite{DBLP:journals/corr/abs-1904-01067}\cite{Gambs:2012:RAT:2352970.2352999}. Either of the two approaches can be used to reconstruct the dataset and it is a one time operation done during the setup phase of the attack as described in Section~\ref{sec:attack}.

\section{Timing Channels In Neural Networks}\label{sec:analysis}

\pgfplotsset{footnotesize,height=5cm,width=0.33\textwidth}

\begin{figure*}[!t]
\begin{center}
\resizebox{2\columnwidth}{!}{%
\begin{tabular}{lllll}

\begin{tikzpicture}

\begin{axis}[
title={(a)},
title style={at={(0.5,0)},anchor=north,yshift=-20},
ylabel={Execution Time},
xlabel={Multiplications},
grid=major
]
\addplot[
    color=black,
    mark=*,
    mark options={solid}
    ]
    coordinates {
	(128*128,0.046)(256*128,0.127)(256*256,0.236)(512*256,0.3749)(512*512,0.661)(1024*512,1.144)(1024*1024,2.534)(2048*1024,4.598)(2048*2048,9.25)(4096*2048,18.98)(4096*4096,38.89)
    };

\end{axis}
\end{tikzpicture}

\begin{tikzpicture}
\begin{axis}[title={(b)},
title style={at={(0.5,0)},anchor=north,yshift=-20},
xlabel={Number of Layers},
ylabel={Execution Time},
grid=major]
\addplot[
    color=black,
    mark=*,
    mark options={solid}
    ]
    coordinates {
	(6,0.0087898489)(11,0.139376951)(16,0.357962286)(18,0.458700957)(21,0.6228808)(24,0.783519824)
    };
\node[label=right:{\scriptsize LeNet}] at (6,0.008) {};
\node[label=right:{\scriptsize AlexNet}] at (11,0.139376951) {};
\node[label=right:{\scriptsize VGG11}] at (16,0.357962286) {};
\node[label=right:{\scriptsize VGG13}] at (18,0.458700957) {};
\node[label=right:{\scriptsize VGG16}] at (20.5,0.6228808) {};
\node[label=right:{\scriptsize VGG19}] at (18,0.783519824) {};
\end{axis}
\end{tikzpicture}

\begin{tikzpicture}
\begin{axis}[title={(c)},
title style={at={(0.5,0)},anchor=north,yshift=-20},
xlabel={Number of Layers},
ylabel={Execution Time},
grid=major]
\addplot[
    color=black,
    mark=*,
    mark options={solid}
    ]
    coordinates {
	(20,0.722430514)(36,1.646826879)(52,1.9543501)(102,3.52180741)(154,5.26268606)(171,5.671749851)
    };
\node[label=right:{\scriptsize Resnet18}] at (21,0.722430514) {};
\node[label=right:{\scriptsize Resnet34}] at (37,1.3) {};
\node[label=right:{\scriptsize Resnet52}] at (53,1.9543501) {};
\node[label=right:{\scriptsize Resnet101}] at (103,3.5) {};
\node[label=right:{\scriptsize Resnet152}] at (130,4.7) {};
\node[label=right:{\scriptsize Resnet161}] at (112,5.8) {};
\end{axis}
\end{tikzpicture}
\end{tabular}
}
\caption{(a) Fully Connected Layer: Total Execution time varies linearly with the total number of multiplications between the weights of the current layer and the input from the previous layer; (b) Simple Neural Networks Topology: Total Execution Time varies linearly with the number of layers of the neural network architecture; (c) Complex Neural Networks Topology: Total Execution Time varies linearly with the number of layers of the neural network with skip connections between different layers.}
\label{fig:arch_skip}
\end{center}
\end{figure*}
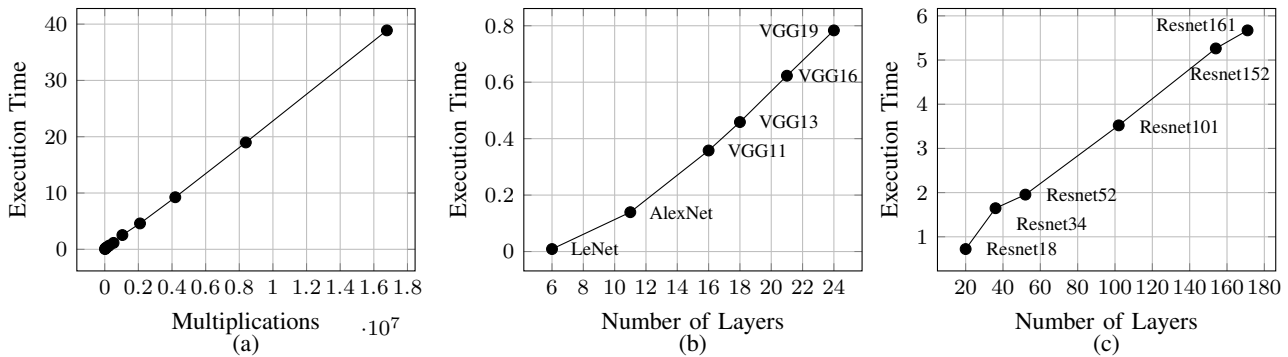

There exists a direct relation between execution time of Neural Networks and their dependence on various hyperparameters for different Neural Network layers as shown in this section.
This dependence of the execution time on the Neural Network hyperparameters allows an adversary to infer the architecture details using the total execution time which forms the basis of the attack.
A typical Convolutional Neural Network has three types of layers based on operations: convolution layer, maxpool layer and fully connected layer and each of the layer has stride, kernel size and number of filter as the hyperparameters.

\textbf{Convolution Layer.} Convolution is a weighted sum operation which computes the multiplication of the parameters ($\theta_{ij}^{(l)}$) of layer $l$ and the input feature map ($z_{ij}^{l-1}$) and adds the results, $x_i^{(l)}=\sum_i \sum_j \theta_{ij}^{(l)} \times z_{ij}^{(l-1)}$. The execution time of convolution layers is proportional to the number of multiplications (shown in Figure~\ref{fig:arch_plain}(a)) which is given by: $o_w \times o_h \times c_o \times f_w \times f_h \times c_i$ where $o_w$ and $o_h$ are the output matrix width and height, $c_i$ and $c_o$ are the input and output number of channels and $f_h$ and $f_w$ represent the filter width and filter height.

\textbf{Maxpool Layer.} Maxpool computes the maximum of the $k \times k$ region of preceding layer feature map where $k$ is the size of the kernel. Hence, the computation of maxpool depends on the number of filters of the feature map and the kernel size ($k$). The execution time increases with increasing filter size and increasing kernel size as shown in Figure~\ref{fig:arch_plain} (b). For stride, there is an inverse relation between the execution time and the stride as shown in Figure~\ref{fig:arch_plain} (c). The output size decreases with increasing stride which results in a decrease in the execution time due to fewer number of multiplications.

\textbf{Fully Connected Layer.} Fully connected layer performs a matrix vector multiplication between the parameters of the Neural Network and the input image map. Formally, the matrix vector multiplication of parameters $\theta_{ij}^{(l)}$ for layer $l$ with the activation of previous layer $z_j^{(l-1)}$ is given by $x_i^{(l)}=\sum_j \theta_{ij}^{(l)} \times z_j^{(l-1)}$. Given two fully connected layers of input nodes $m$ and output nodes $n$, the execution time varies linearly with the total number of multiplications, $m \times n$ and the linear relation can be seen in Figure~\ref{fig:arch_skip} (a).

\textbf{Variation with Depth.} Neural Networks are embarrassingly parallel and all the computations in one layer can be executed in parallel to some extent which enables optimisations like model and data parallelism to improve the performance \cite{NIPS2012_4824}.
However, due to sequential computation of Neural Networks, the total execution time is the sum of the execution time of individual layers.
In Figure~\ref{fig:arch_skip} (b) and Figure~\ref{fig:arch_skip} (c), the execution time increases linearly with the increase in the network architecture depth for both simple and complex topologies.
The simple topologies of Neural Networks include LeNet, AlexNet and VGG architectures while the complex topologies include Resnet architectures which have skip connections between the layers.
This linear relation between the number of layers and the total execution time forms the basis of the attack methodology to infer the depth of the Neural Network given the execution time.

\textbf{Hardware Factors.} While the Neural Networks fetch the data and parameters from the memory, the program could incur latencies during reading or storing data, contention of threads and inefficient caching.
Since the hardware used for all the models is same, these factors are assumed to effect the execution time of all the Neural Networks in the same manner.

\section{Attack Methodology}\label{sec:attack}

Side channels reveal only a part of the secret in the target system and identifying the rest of the secret is modelled as a search problem \cite{yan2018cache}.
The proposed attack is broadly divided into three phases as shown in Figure~\ref{overview}:
\begin{itemize}
\item \textbf{Setup Phase}: Adversary aggregates the dataset by measuring the execution time of multiple models with different hyperparameters, on a particular hardware, to be used in the actual attack. Further, the adversary reconstructs the training dataset using iterative membership inference \cite{salem2018ml} or dataset reconstruction attack \cite{DBLP:journals/corr/abs-1904-01067}. This is a one time operation required to be performed prior to the attack.
\item \textbf{Attack Phase}: Adversary queries the target Neural Network and measures the total execution time averaged over all queries. A regressor is trained on the attacker dataset which is used to infer the target Neural Network depth.
\item \textbf{Reconstruction Phase}: Adversary searches for an optimal Neural Network with test accuracy close to that of the target Neural Network model within the reduced search space by making the depth of the Neural Network constant as inferred from the attack.
\end{itemize}

\subsection{Setup Phase}

The adversary during the setup phase reconstructs the training data with the prediction of target model ($f_{target_i}(x)$) as labels instead of true labels for model distillation and creates a dataset containing the execution time of Neural Network architectures with varying hyperparameters.
This a one time operation required prior to performing the attack.

\textbf{Reconstructing Training Data.} Since, the attack assumes a weak adversary with no knowledge about the training data, the adversary needs to identify and generate data samples used as the training dataset.
The underlying data distribution is assumed to be known, the adversary can sample data points $x$ from the underlying data distribution.
The adversary queries the target model with data sample $x$, and based on the target model output posterior, determines if the data sample belongs to the training data $D_T$ of the target Neural Network or not, by selecting a suitable threshold \cite{salem2018ml}.
Given the knowledge of the data samples used as part of the target model training dataset, the adversary labels the input instances ($x$) with the predictions of the target model ($f_{target}(x)$) instead of the true label ($y$). The aggregated training data $(x_i,f_{target}(x_i))$ includes soft target model predictions ($f_{target}(x_i)$) instead of hard labels ($y$), which ensures that the substitute model learns to mimic the functionality of the target model \cite{hinton2015distilling}\cite{DBLP:journals/corr/BaC13}.

\textbf{Creating Attack Model Timing Dataset.} The adversary populates the attack model dataset which contains the time taken for inference and corresponding depth of the Neural Networks along with the number of parameters for the corresponding model as shown in Table~\ref{table2}.
Formally, for different model architectures and depth ($K$), the attacker collects the corresponding execution time ($T$) to generate the attacker dataset $D_A$ = \{$(T_1 ,K_1 ), \cdots ,(T_N ,K_N )\}$. This dataset is specific to a particular hardware and the collected data can then be used for stealing any model run on the same hardware.

\begin{table}[htb!]
\centering
\begin{tabular}{|c|c|c|}
\hline
\textbf{Architecture} & \textbf{Parameters} & \textbf{Inference Time (s)}\\
\hline
\multirow{3}{*}{VGG16} & 138,357,544 & 0.59683408 \\
& 156,053,800 & 0.86338694\\
& 133,048,360 & 0.49502961 \\
\hline
\multirow{3}{*}{VGG19} & 143,667,240  & 0.7642632 \\
& 168,441,896 & 1.11189311 \\
& 136,588,328 & 0.55131464\\
\hline
\end{tabular}
\caption{\textbf{Sample of Adversary's Dataset.} Depth of the architecture and the number of parameters as input features for the regressor.}
\label{table2}
\end{table}

\subsection{Attack Phase}

\textbf{Query.} The adversary sends queries to the target model and computes the overall execution time averaged across all the queries.
For each additional query of model extraction attacks proposed previously, a new bit of information is leaked which reduces the entropy of the target black box model.
However, this requires a large number of queries due to the large number of parameters of deep Neural Network architectures.
In the proposed attack, each of the query to the target model reveals the same bit of information (execution time averaged across all queries) which allows to make constant number of queries, independent of the architecture.

\textbf{Regression.} The average execution time measured from the target Neural Network ($t$) is used to estimate the depth of Neural Network ($k$) using regressor trained on the attack model dataset created during the setup phase, $R(k) = E\{K \vert T = t\}$.
For current experiments, the attacker dataset for training the regressor model is created using 100 Neural Networks with different depth and parameters as shown in Figure~\ref{scatter}.

\pgfplotsset{footnotesize,height=5.3cm,width=0.8\columnwidth}
\begin{figure}[htb!]
\begin{center}
\begin{tikzpicture}
\begin{axis}[%
xlabel={Execution Time(s)}, ylabel={Number of Layers}, grid=major,
scatter/classes={%
    a={mark=*,mark options={solid},draw=black}}]
\addplot[scatter,only marks,%
    scatter src=explicit symbolic]%
table[meta=label] {
x y label
0.7642632 24 a
0.674304789 22 a
0.55920943 22 a
0.714724269 23 a
0.619721165 23 a
0.58131464 24 a
0.716728751 26 a
0.651927690 25 a
0.72859727 27 a
0.57450162 21 a
0.48480499 21 a
0.35156030 16 a
0.487170366 16 a
0.2486484129 16 a
0.43352229 18 a
0.532749272 18 a
0.36160554 18 a
0.401576198 17 a
0.49417733 17 a
0.340565053 17 a
0.49747443 19 a
0.417900943 19 a
0.294650654 15 a
0.4527853 15 a
0.2412370 15 a
0.236719541 14 a
0.327085508 14 a
0.207642712 14 a
0.185943815 13 a
0.25997309 13 a
0.13972117 13 a
0.015124014 6 a
0.017141882 6 a
0.016748258 7 a
0.0188383 7 a
0.0155131 7 a
0.026086211 9 a
0.02216158 8 a
0.0150779 8 a
0.01811455 8 a
0.01320973 5 a
0.142865239 11 a
0.114419951 10 a
0.08470160 10 a
0.13226054 10 a
0.098939943 9 a
0.07593683 9 a
0.1116653 9 a
0.084226102 8 a
0.0526757 8 a
0.098145046 8 a
0.10738156 11 a
0.15729382 12 a
0.24656767 12 a
0.170794195 12 a
0.21973929 13 a
0.29143093 13 a
0.1763891 13 a
0.259405379 14 a
0.28576108 14 a
0.212981777 14 a
0.183592753 11 a
0.44710270 18 a
0.604612316 18 a
0.39547864 18 a
0.5193142 19 a
0.67968742 19 a
0.448244426 19 a
0.510759344 20 a
0.653228242 20 a
0.479291455 20 a
0.625086684 23 a
0.48677991 23 a
0.409641069 17 a
0.57591108 17 a
0.336897214 17 a
0.363749254 16 a
0.28077226 16 a
0.5151859 16 a
0.2811809549 15 a
0.36787389 15 a
0.243812743 15 a
0.57683408 21 a
0.49502961 21 a
0.61284602 22 a
0.568350783 22 a
0.652385803 23 a
0.582251877 23 a
0.724749135 24 a
0.626187309 24 a
0.547266757 20 a
0.615754716 20 a
0.481281377 20 a
0.495092882 19 a
0.547714346 19 a
0.4029508719 19 a
0.420202944 18 a
0.601368664 18 a
0.39523135 18 a
};
\end{axis}
\end{tikzpicture}
\caption{\textbf{Scatter Plot of the Adversary's Dataset.} Execution time vs Number of layers. Each point in the plot is obtained by varying the parameters of the neural network layers and the number of layers in the neural network. The attacker fits a regressor to predict the depth of the neural network given the execution time.}
\label{scatter}
\end{center}
\end{figure}
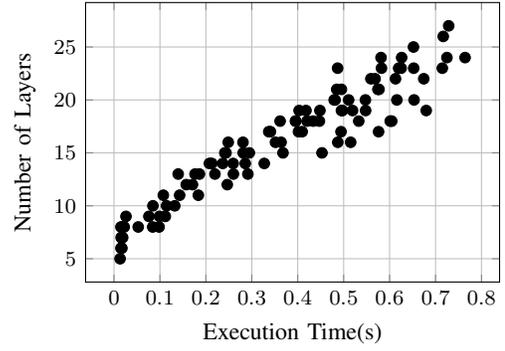

\subsection{Reconstruction Phase}

The search space of all possible Neural Networks is very large and complex due to which designing the Neural Network manually is hard and requires optimal search strategies to reduce and simplify the search space.
The depth of the Neural Network reduces the search space and the model exploration is automated using Reinforcement Learning which outputs the best model architecture in the constrained search space \cite{DBLP:journals/corr/ZophL16}.
A Recurrent Neural Network (RNN) based controller predicts the hyperparameters of each layer in the template Neural Network sampled using the reward from the previous proposed architecture.
The parameters of controller RNN $\theta_c$ are then optimised based on performance of the predicted architecture using policy gradient method.

Formally, the controller predicts the architectures through actions $a_{1:T}$ and the predicted model tries to achieve accuracy $R$ which is used to compute the reward signal to train the controller.
The policy function maps the proposed architecture to a real number (accuracy) which is used to compute the reward for the RNN controller. The policy weights are optimised to ensure that newer architectures proposed by the RNN controller have a higher performance.
After training the RNN controller for multiple iterations, the proposed substitute architecture is optimal, i.e, has the highest expected accuracy among all the models within the search space.
The goal is to maximise the expected accuracy of sampled architecture given by,
\begin{equation}
J(\theta_c) = E_{P(a_{1:T};\theta_c)}[R]
\end{equation}

while minimising the difference in test accuracy between the target and proposed substitute model, $R^{target}_{test} - R^{substitute}_{test}$.

\pgfdeclarelayer{background}
\pgfdeclarelayer{foreground}
\pgfsetlayers{background,main,foreground}

\tikzstyle{input} = [rectangle, rounded corners, minimum width=0.15cm, minimum height=1.5cm,text centered, draw=black, fill=gray!15]
\tikzstyle{layer1} = [rectangle, rounded corners, minimum width=0.25cm, minimum height=1.5cm,text centered, draw=black, fill=gray!15]
\tikzstyle{layer2} = [rectangle, rounded corners, minimum width=0.25cm, minimum height=1cm,text centered, draw=black, fill=gray!15]
\tikzstyle{layer3} = [rectangle, rounded corners, minimum width=0.25cm, minimum height=0.6cm,text centered, draw=black, fill=gray!15]
\tikzstyle{layer4} = [rectangle, rounded corners, minimum width=0.25cm, minimum height=0.25cm,text centered, draw=black, fill=gray!15]
\tikzstyle{write} = [rectangle, rounded corners, minimum width=2.5cm, minimum height=1.5cm,text centered, draw=black, fill=gray!15]
\tikzstyle{discriminator} = [rectangle, rounded corners, minimum width=3cm, minimum height=1.2cm,text centered, draw=black, fill=gray!15]

\begin{figure}[htb!]
\centering
\resizebox{.5\textwidth}{!}{
\begin{tikzpicture}[node distance=2cm, line width=1pt,every node/.style={align=center}]

\node (en1) [layer1]  at (-4,0) {};
\node (en2) [layer2]  at (-3.5,0) {};
\node (en3) [layer3]  at (-3,0) {};
\node (en4) [layer4]  at (-2.5,0) {};
\node (pred) [input]  at (-1.75,0) {};

\begin{scope}[on background layer]
    \node (target) [fit=(en1) (en2) (en3) (en4) (pred), fill= gray!30, rounded corners, inner sep=.2cm, label={above:Target Model}] {};
\end{scope}

\node (en11) [layer1]  at (-4,-3) {};
\node (en12) [layer2]  at (-3.5,-3) {};
\node (en13) [layer3]  at (-3,-3) {};
\node (en14) [layer4]  at (-2.5,-3) {};
\node (pred1) [input]  at (-1.75,-3) {};

\begin{scope}[on background layer]
    \node (substitute) [fit=(en11) (en12) (en13) (en14) (pred1), fill= gray!30, rounded corners, inner sep=.2cm, label={below:Substitute Model}] {};
\end{scope}

\node (controller) [discriminator] at (-7,-3) {RNN Controller};

\node (loss) [draw, label=right:\LARGE \ding{203}] at (1.25,-1.5) {Loss \\$L(y^{substitute},y^{target})$};

\draw[thick,->] (en4.east) -- node[anchor=north,xshift=1.5cm,] {$y^{target}$} (pred.west);
\draw[thick,->] (en14.east) -- node[anchor=north,xshift=1.5cm,] {$y^{substitute}$} (pred1.west);
\draw[thick,->] (controller.east) -- node[anchor=north] {\LARGE \ding{202}} (substitute.west);
\draw[thick,->] (target.east) -| node[anchor=north] {} (loss.north);
\draw[thick,->] (substitute.east) -| node[anchor=north] {} (loss.south);
\draw[thick,->] (loss.west) -| node[anchor=north,xshift=1.5cm, yshift=0.75cm,] {\LARGE \ding{204}} (controller.north);

\end{tikzpicture}
}
\caption{\textbf{Reinforcement Learning Based Architecture Search with Distillation.} 1) The Recurrent Neural Network controller proposes a Neural Network architecture by selecting values from the state search space based on the the reward from the previous iteration; 2) The proposed model trains by using the predictions of the target model as the output predictions and mimics the behaviour of the target model \cite{hinton2015distilling}\cite{DBLP:journals/corr/BaC13}; 3) The loss of the template and target model predictions is used to compute a reward which is sent to the controller to predict a better performing model}
\label{fig:nas}
\end{figure}
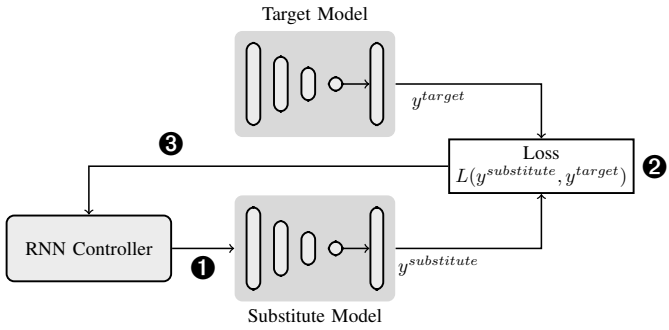

\begin{table*}[hb!]
\centering
\resizebox{\textwidth}{!}{%
\begin{tabular}{|c|c c c c c|}
\hline
\textbf{} & \textbf{Architecture} & \textbf{Parameters} & \textbf{Execution Time} & \textbf{Predicted Depth} & \textbf{True Depth}\\
\hline
\textbf{Model 1} & [32(3), 32(3), MP, 64(3), 64(3), MP, 128(3), 128(3), MP] & 309,290 & 0.036057 & 8.8 (RF); 8.1 (BDT) & 9\\
\textbf{Model 2} & [32(3), 32(3), MP, 64(3), 64(3), MP, 128(3), 128(3), MP, 256(3), MP] & 595,242 & 0.10738 & 10.15 (RF); 10.02 (BDT)& 11\\
\textbf{Model 3} & [32(3), 32(3), MP, 64(3), 64(3), MP, 128(3), 128(3), 128(3), MP, 256(3), 256(3), MP] & 1,334,442 & 0.18594 & 12.8 (RF); 12.6 (BDT) & 13\\
\hline
\end{tabular}
}
\caption{\textbf{Regression Model Predictions.} The estimated depth of the neural network is rounded to nearest larger integer. The depth is correctly estimated using ensemble regressors: Random Forest (RF) and Boosted Decision Trees (BDT). The architecture of the model is in format: filters (kernel size) and MP is the Maxpool Layer}
\label{table7}
\end{table*}

\begin{table*}[hb!]
\centering
\resizebox{\textwidth}{!}{%
\begin{tabular}{|c|c c c c|}
\hline
\textbf{} & \textbf{Reconstructed Architecture} & \textbf{Parameters} & \textbf{Original Accuracy} & \textbf{Reconstruction Accuracy}\\
\hline
\textbf{Model 1} & [64(3), 32(5), 128(3), 64(5), 128(3), 64(5), 32(5), 128(3), GAP] & 535,114 & 88.03\% & 86.06\% \\
\textbf{Model 2} & [32(5), 32(5), 64(3), 32(3), 64(5), 128(3), 128(3), 32(3), 64(3), 64(5), GAP] & 639,978 & 89.26\% & 85.65\% \\
\textbf{Model 3} & [64(3), 128(3), 128(3), 32(5), 64(5), 128(3), 128(3), 32(3), 128(5), 128(3), 64(5), 128(5), GAP] & 889,834 & 90.19\% & 85.3\%\\
\hline
\end{tabular}
}
\caption{\textbf{Evaluation of Reconstructed Models using Reinforcement Learning.} The test accuracy of the reconstructed model is close to the original accuracy of the target model. The architecture of the model is in format: filters (kernel size) and GAP is the Global Average Pooling}
\label{table8}
\end{table*}

Since the reward signal $R$ is non-differentiable, we need to use a policy gradient method to iteratively update the parameters of the RNN controller $\theta_c$.
This work uses the $REINFORCE$ algorithm to directly update the policy weights using stochastic gradient ascent and improve the policy \cite{Williams1992}:
\begin{equation}
\bigtriangledown_{\theta_c} J(\theta_c) = \sum_{t=1}^{T} E_{P(a_{1:T};\theta_c)}\big[\bigtriangledown_{\theta_c} \log P(a_t|a_{(t-1):1};\theta_c)R\big]
\end{equation}

The training of each proposed model is done using knowledge distillation where the loss function for training the substitute model is the distillation L2 loss function \cite{hinton2015distilling} between the substitute model predictions ($y^{template}$) and target model($y^{target}$) predictions instead of the true labels($y$) for a given data point $x$ $\epsilon$ $D$ and is given as,
\begin{equation}
L=\sum_{i=1}^{n}\left ( y^{target}_i- y^{template}_i \right )^2
\end{equation}
For each iteration of training the substitute model, the substitute model learns to mimic the predictions of the target model making the test performance similar (Figure~\ref{fig:nas}).

\section{Evaluation}\label{sec:evaluation}

\textbf{Data.} For all the experiments, CIFAR10 dataset \cite{cifar10} is used which contains 60,000 32x32 colour images in 10 classes with around 6000 images per class and the classes are mutually exclusive. For training, 50,000 images are used while 10,000 images are used for testing.

\textbf{Experiment Setup.} The processor used for evaluation and experimentation is Intel Xeon Gold 5115 server processor with a 2.4GHz clock speed, 196GB of main memory and 40 cores. All the reported number are an average of 20 inference runs. Accurate timing of the Neural Network inference is done using $process\_time()$ from the $time()$ python library which computes the time interval for running the inference using the CPU counter and is not effected by execution of other unrelated processes. The clock has a tick rate (ticks/s) of 10,000,000 which indicates a high resolution of 1e-07.

\subsection{Regression}

Ideally, on executing a Neural Network on a hardware accelerator, the total execution time depends solely on the number of layers due to their sequential computation.
Hence, varying the number of neurons or filters in a particular layer in a deep Neural Network should not change the overall execution time as individual computations within the same layer can be performed in parallel.
This property of Neural Networks makes them embarrassingly parallel \cite{DBLP:journals/corr/SzeCYE17}.
However, in practice, a variation in number of parameters within a particular layer shows a deviation in execution time due to inefficient parallelism as seen in Figure~\ref{scatter}.

To address this, it is important to train a good regressor which captures the maximum variance in the timing dataset.
A good regressor should be able to capture and explain the variance of the dataset through its predictions and generalize well over the data samples.
The evaluation of five different regressor models on the attacker dataset is done using $R^2$ score metric to measure the variance explained by the model and the mean squared error (MSE) to measure the error in estimating the depth of the network.

Ridge and Support Vector regressors are linear models, while Decision Tree Regressor builds a tree structure over different features in the dataset by minimising the standard deviation.
Boosting is an ensemble technique to combine multiple sequential decision trees and fit the data, while improving the error from the pervious model.
On the other hand, Random Forest regressor fits subsets of data over multiple decision trees independently and outputs the mean of prediction from each of the model.

Based on the results, ensemble approaches like boosted decision trees (BDT) and random forrest (RF) regressor have a higher $R^2$ score and lower mean squared error to estimate the depth more accurately as compared to the linear models which fail to capture the variance of the attacker dataset (Table~\ref{table3}).
The estimated depth of the ensemble regressors on the target deep Neural Networks from the corresponding execution time on VGG based target Neural Network architectures is shown in Table~\ref{table7}.
Since the output of the regressor is a continuous variable, the prediction of the regressor is rounded to the nearest larger integer. For all the three Neural Networks used for evaluation, the regressors estimate the correct depth from the total execution time.

\begin{table}[h]
\centering
\begin{tabular}{|c|c c|}
\hline
\textbf{Regressor} & \textbf{Mean Squared Error} & \textbf{$R^2$ Score}\\
\hline
\textbf{Support Vector} & 7.5405  & 0.7295 \\
\textbf{Decision Tree} & 5.375 & 0.80719 \\
\textbf{Linear (Ridge)} & 4.46533 & 0.8398 \\
\textbf{Boosted Decision Tree} & 4.1947 & 0.8495 \\
\textbf{Random Forrest} & 3.7664 & 0.8648 \\
\hline
\end{tabular}
\caption{\textbf{Evaluation of Regression Models using $R^2$ score and Mean Squared Error (MSE).} Ensemble approaches estimate the depth better with higher $R^2$ score and lower MSE.}
\label{table3}
\end{table}

\subsection{Reconstruction using Reinforcement Learning}

Once the adversary has estimated the depth of the Neural Network, the information is used to reduce the search space. Now, the adversary has to search for the optimal substitute Neural Network with test accuracy close to the target Neural Network.

The Reinforcement Learning based architecture search is evaluated by fixing the depth of the model architectures inferred using the regressor.
The search is further constrained by specifying the convolutional layer parameter range for the kernel size ($k$) $\epsilon$ \{3,5\} and the number of filter ($n_f$) $\epsilon$ \{32,64,128\} which are commonly used hyperparameters values used in all state of the art networks \cite{DBLP:journals/corr/SimonyanZ14a}.
The architecture search approach explores the space of 50 models, and outputs the architecture corresponding the highest accuracy.
To improve the performance of the substitute model, we use fully convolutional net architecture by replacing maxpool layer with convolutional layers with higher stride \cite{DBLP:journals/corr/SpringenbergDBR14}. The reward used for updating the controller is the maximum validation accuracy of the last 5 epochs cubed which is clipped in the range (-0.05, 0.05) to ensure that the gradients do not overshoot.
The controller RNN includes 1 LSTM cell and 32 hidden units and each proposed template model is trained for 20 epochs.
For all the three model, the test accuracy of the substitute model generated is within 5\% of the target model architecture as shown in Table~\ref{table8}.

\section{Mitigation}\label{sec:mitigation}

The main reason for the manifestation of timing side channels in Neural Networks is the sequential computation of layers which determine the total execution time.
It is important to design Neural Networks resistant to timing side channels to prevent model extraction. Some of the possible defences are discussed in this section.

\textbf{Adding Noise to Execution Time.} Instead of having a Neural Network with timing dependent on the depth of the Neural Network, one defence mechanism is to design Neural Networks without the dependency of the execution time on the number of layers and hyperparameters.
Additional noise to the total execution time can be added in the form of latency by including dummy computations and layers.
However, this results in a model security and utility tradeoff which is a concern in real time critical applications where the performance of the model within time constraints is vital.

\textbf{Adversarial Machine Learning.} The second mechanism is a training phase defence where the attack and defence game can be viewed as an adversarial machine learning problem.
The goal of the attacker is to fit the best possible curve or function to the attacker dataset for regression while the goal of the defender is to poison the dataset with wrong data instances such that the regressor makes wrong predictions.
The defender injects adversarial examples with incorrect timing and depth values resulting in incorrect regressor estimation.

\section{Discussion}\label{sec:discussion}

\textbf{Variation Across Datasets.} The timing distribution for Neural Networks is specific to the training data used. For the same architecture, using different datasets results in different execution time due to different number of computations as shown in Table~\ref{table4}.

\begin{table}[htb!]
\centering
\begin{tabular}{|c|c|c|}
\hline
\textbf{Architecture} & \textbf{Dataset} & \textbf{Inference Time (s)}\\
\hline
\multirow{2}{*}{Alexnet}
& MNIST & 0.28958\\
& CIFAR10 & 0.36527 \\
\hline
\multirow{2}{*}{VGG}
& MNIST & 0.44178 \\
& CIFAR10 & 0.63833\\
\hline
\end{tabular}
\caption{Same architecture trained on different datasets have different timing distribution.}
\label{table4}
\end{table}

Different dataset have images of different sizes and properties due to which the intermediate input feature maps require different number of multiplications.
For instance, MNIST dataset has images of size 28 $\times$ 28 $\times$ 1, while CIFAR10 dataset has images of size 32 $\times$ 32 $\times$ 3 which results in difference in number of computations to be performed.
Hence, the regressor is specific to a particular timing distribution unique to a dataset and a different attack model has to be trained to fit different timing distribution.

\textbf{Extending to Remote Setting.} While the evaluation the attack is on a local model, this can be extended to remote setting like in MLaaS where the model is deployed on a Cloud server.
The total round trip time in case of remote setting also includes some additional noise during propagation in the form of jitter as well as the time taken for network propagation \cite{Crosby:2009:OLR:1455526.1455530}.
A round trip time model for remote timing attacks is given below where the total response time ($t_{res}$) is a linear function of the scaled processing time ($t_{proc}$), network propagation time ($t_{net}$) and the jitter.

\begin{equation}
t_{res} = a \times t_{proc} + t_{net} + jitter
\end{equation}

To extract the processing time of the Neural Network from the total round trip time, the additional network time and jitter have to be estimated and filter them from the round trip time.

\textbf{Model Extraction Defences.} Several defences have been proposed to mitigate the attacks that exploit the information from output predictions.
Suppressing the information provided by output logits reduces the accuracy of the substitute model but degrades the utility \cite{tramer2016stealing}.
Stateful defence mechanism to monitor and detect a variation in the input query distribution \cite{DBLP:journals/corr/abs-1805-02628} or raise an alert if the information gained by an adversary exceeds a threshold \cite{kesarwani2017model}.
Trusted hardware like Intel SGX can protect the confidentiality and integrity of the model by moving the model offline to the user's system \cite{hanzlik2018mlcapsule}.
For attacks that rely on memory access patterns, implementation using Oblivious RAM could help to hide the access pattern \cite{Stefanov:2013:POE:2508859.2516660}.
All the defences mentioned are proposed for attacks that use the output prediction scores but none of these approaches can help to mitigate timing side channels.

\section{Related Work} \label{sec:related}

Deep Neural Network attributes can be extracted from the input-output relationship of the target network \cite{inproceedings}\cite{orekondy19knockoff}.
For example, given a black box neural network, an adversary queries the model with data instances $x$ to obtain corresponding output predictions $f(x)$.
These input-output pair ($x$,$f(x)$) is used for training the substitute model, whose parameters and hyperparameters can be computed by solving a system of linear equations between the input, output predictions and the unknown parameters \cite{tramer2016stealing}.
Further, other hyperparameters of the objective function can be solved by finding an approximate solution to a system of overdetermined equations \cite{wang2018stealing}.
Unfortunately, such attacks rely on large number queries since each new query provides the solution for a different unknown variable.
Further, for very deep neural network architectures with large number of layers and millions of parameters, the computation cost is high.
Unlike these attacks, timing side channels can be exploited in constant number of queries since each query reveals the same model attribute and the approach is scalable to deep networks with millions of parameters.

Several machine learning models trained to predict the model attributes based on the input output relationships can infer significant number of model attributes \cite{oh2018towards}.
However, such techniques incur a high computational cost for training large number of machine learning attack models.
For a simple digit classification task, it takes 10k attack models trained over 40 GPU days.

Alternatively, Side Channel leakage during the model execution provides fine-grained information about the target model in the form of cache misses, memory access pattern \cite{Naghibijouybari:2018:RIG:3243734.3243831}\cite{DBLP:journals/corr/abs-1903-03916}, power consumption profile \cite{cryptoeprint:2018:477} and hardware performance counters \cite{Alam2018HowSA}.
The read-after-write dependencies for the inputs activation filters and output activation filters are different which reveal the dimensions and type of individual layers \cite{Hua:2018:REC:3195970.3196105}.
Assuming shared resources between the target model (victim)process and the attacker process, an adversary can monitor the number of calls, the size of matrix dimensions to identify the number of layers and hyperparameter details in the Neural Network \cite{hong2018security}.
Further, cache attacks can distinguish different activation function like relu, sigmoid and tanh by monitoring the probe addresses \cite{yan2018cache}.
Given the power traces during the execution of models, algorithms like differential power analysis, correlated power analysis and horizontal power analysis can be used to extract the the number of parameters in each layers, values of each parameters, total number of layers and the type of activation function \cite{cryptoeprint:2018:477}.

However, these side channel attacks either assume a strong adversary with physical access to the hardware or require shared resources between the processes.
Unlike these side channel attacks, the proposed approach considers a weak adversary with (remote) blackbox access to the target model.

\section{Conclusions}\label{sec:conclusion}

This paper shows that Neural Networks are vulnerable to timing side channels attacks as the total execution time depends on the sequential computation along the number of layers or depth.
For a weak adversary in a black box setting, the timing channel vulnerability can be exploited to infer the depth of the Neural Network architecture.
The evaluation of various regressors on the timing data shows that the ensemble based regressors perform better than their linear counterparts based on the $R^2$ score and Mean Score Error values.
Further, the search problem of extracting a Neural Network architecture by exploiting side channels can be addressed efficiently using Reinforcement Learning.
This approach can be used with other attacks like cache attacks and memory access pattern monitoring to accurately identify the substitute model close to the target model.
The attack is evaluated on VGG like deep learning architectures and it is shown that a substitute model can be reconstructed within 5\% of the test accuracy of the target Neural Network.


{\footnotesize
\bibliographystyle{IEEEtranS}
\bibliography{paper.bib}

\begin{thebibliography}{10}
\providecommand{\url}[1]{#1}
\csname url@samestyle\endcsname
\providecommand{\newblock}{\relax}
\providecommand{\bibinfo}[2]{#2}
\providecommand{\BIBentrySTDinterwordspacing}{\spaceskip=0pt\relax}
\providecommand{\BIBentryALTinterwordstretchfactor}{4}
\providecommand{\BIBentryALTinterwordspacing}{\spaceskip=\fontdimen2\font plus
\BIBentryALTinterwordstretchfactor\fontdimen3\font minus
  \fontdimen4\font\relax}
\providecommand{\BIBforeignlanguage}[2]{{%
\expandafter\ifx\csname l@#1\endcsname\relax
\typeout{** WARNING: IEEEtranS.bst: No hyphenation pattern has been}%
\typeout{** loaded for the language `#1'. Using the pattern for}%
\typeout{** the default language instead.}%
\else
\language=\csname l@#1\endcsname
\fi
#2}}
\providecommand{\BIBdecl}{\relax}
\BIBdecl

\bibitem{amazonec2}
``Amazon ec2 instance types,''
  \url{https://aws.amazon.com/ec2/instance-types/}, 2018.

\bibitem{amazonsage}
``Amazon sagemaker instance types,''
  \url{https://aws.amazon.com/sagemaker/pricing/instance-types/}, 2018.

\bibitem{cryptoeprint:2018:477}
\BIBentryALTinterwordspacing
``{CSI} {NN}: Reverse engineering of neural network architectures through
  electromagnetic side channel,'' in \emph{28th {USENIX} Security Symposium
  ({USENIX} Security 19)}.\hskip 1em plus 0.5em minus 0.4em\relax Santa Clara,
  CA: {USENIX} Association, 2019. [Online]. Available:
  \url{https://www.usenix.org/conference/usenixsecurity19/presentation/batina}
\BIBentrySTDinterwordspacing

\bibitem{Alam2018HowSA}
M.~Alam and D.~Mukhopadhyay, ``How secure are deep learning algorithms from
  side-channel based reverse engineering?'' in \emph{DAC}, 2018.

\bibitem{DBLP:journals/corr/BaC13}
\BIBentryALTinterwordspacing
J.~Ba and R.~Caruana, ``Do deep nets really need to be deep?'' in
  \emph{Advances in Neural Information Processing Systems 27}, Z.~Ghahramani,
  M.~Welling, C.~Cortes, N.~D. Lawrence, and K.~Q. Weinberger, Eds.\hskip 1em
  plus 0.5em minus 0.4em\relax Curran Associates, Inc., 2014, pp. 2654--2662.
  [Online]. Available:
  \url{http://papers.nips.cc/paper/5484-do-deep-nets-really-need-to-be-deep.pdf}
\BIBentrySTDinterwordspacing

\bibitem{bishop2006pattern}
C.~M. Bishop, \emph{Pattern Recognition and Machine Learning (Information
  Science and Statistics)}.\hskip 1em plus 0.5em minus 0.4em\relax Berlin,
  Heidelberg: Springer-Verlag, 2006.

\bibitem{Crosby:2009:OLR:1455526.1455530}
\BIBentryALTinterwordspacing
S.~A. Crosby, D.~S. Wallach, and R.~H. Riedi, ``Opportunities and limits of
  remote timing attacks,'' \emph{ACM Trans. Inf. Syst. Secur.}, vol.~12, no.~3,
  pp. 17:1--17:29, Jan. 2009. [Online]. Available:
  \url{http://doi.acm.org/10.1145/1455526.1455530}
\BIBentrySTDinterwordspacing

\bibitem{DSJ12371}
\BIBentryALTinterwordspacing
V.~Duddu, ``A survey of adversarial machine learning in cyber warfare,''
  \emph{Defence Science Journal}, vol.~68, no.~4, pp. 356--366, 2018. [Online].
  Available:
  \url{https://publications.drdo.gov.in/ojs/index.php/dsj/article/view/12371}
\BIBentrySTDinterwordspacing

\bibitem{fredrikson2015model}
M.~Fredrikson, S.~Jha, and T.~Ristenpart, ``Model inversion attacks that
  exploit confidence information and basic countermeasures,'' in
  \emph{Proceedings of the 22nd ACM SIGSAC Conference on Computer and
  Communications Security}.\hskip 1em plus 0.5em minus 0.4em\relax ACM, 2015.

\bibitem{Gambs:2012:RAT:2352970.2352999}
\BIBentryALTinterwordspacing
S.~Gambs, A.~Gmati, and M.~Hurfin, ``Reconstruction attack through classifier
  analysis,'' in \emph{Proceedings of the 26th Annual IFIP WG 11.3 Conference
  on Data and Applications Security and Privacy}, ser. DBSec'12.\hskip 1em plus
  0.5em minus 0.4em\relax Berlin, Heidelberg: Springer-Verlag, 2012, pp.
  274--281. [Online]. Available:
  \url{http://dx.doi.org/10.1007/978-3-642-31540-4_21}
\BIBentrySTDinterwordspacing

\bibitem{hanzlik2018mlcapsule}
L.~Hanzlik, Y.~Zhang, K.~Grosse, A.~Salem, M.~Augustin, M.~Backes, and
  M.~Fritz, ``Mlcapsule: Guarded offline deployment of machine learning as a
  service,'' \emph{arXiv preprint arXiv:1808.00590}, 2018.

\bibitem{hazelwood2018applied}
K.~Hazelwood, S.~Bird, D.~Brooks, S.~Chintala, U.~Diril, D.~Dzhulgakov,
  M.~Fawzy, B.~Jia, Y.~Jia, A.~Kalro \emph{et~al.}, ``Applied machine learning
  at facebook: A datacenter infrastructure perspective,'' in \emph{High
  Performance Computer Architecture (HPCA), 2018 IEEE International Symposium
  on}.\hskip 1em plus 0.5em minus 0.4em\relax IEEE, 2018, pp. 620--629.

\bibitem{hinton2015distilling}
\BIBentryALTinterwordspacing
G.~Hinton, O.~Vinyals, and J.~Dean, ``Distilling the knowledge in a neural
  network,'' in \emph{NIPS Deep Learning and Representation Learning Workshop},
  2015. [Online]. Available: \url{http://arxiv.org/abs/1503.02531}
\BIBentrySTDinterwordspacing

\bibitem{hitaj2017deep}
B.~Hitaj, G.~Ateniese, and F.~P{\'e}rez-Cruz, ``Deep models under the gan:
  information leakage from collaborative deep learning,'' in \emph{Proceedings
  of the 2017 ACM SIGSAC Conference on Computer and Communications
  Security}.\hskip 1em plus 0.5em minus 0.4em\relax ACM, 2017.

\bibitem{hong2018security}
S.~Hong, M.~Davinroy, Y.~Kaya, S.~N. Locke, I.~Rackow, K.~Kulda,
  D.~Dachman-Soled, and T.~Dumitra{\c{s}}, ``Security analysis of deep neural
  networks operating in the presence of cache side-channel attacks,''
  \emph{arXiv preprint arXiv:1810.03487}, 2018.

\bibitem{DBLP:journals/corr/abs-1903-03916}
\BIBentryALTinterwordspacing
X.~Hu, L.~Liang, L.~Deng, S.~Li, X.~Xie, Y.~Ji, Y.~Ding, C.~Liu, T.~Sherwood,
  and Y.~Xie, ``Neural network model extraction attacks in edge devices by
  hearing architectural hints,'' \emph{CoRR}, vol. abs/1903.03916, 2019.
  [Online]. Available: \url{http://arxiv.org/abs/1903.03916}
\BIBentrySTDinterwordspacing

\bibitem{Hua:2018:REC:3195970.3196105}
\BIBentryALTinterwordspacing
W.~Hua, Z.~Zhang, and G.~E. Suh, ``Reverse engineering convolutional neural
  networks through side-channel information leaks,'' in \emph{Proceedings of
  the 55th Annual Design Automation Conference}, ser. DAC '18.\hskip 1em plus
  0.5em minus 0.4em\relax New York, NY, USA: ACM, 2018, pp. 4:1--4:6. [Online].
  Available: \url{http://doi.acm.org/10.1145/3195970.3196105}
\BIBentrySTDinterwordspacing

\bibitem{DBLP:journals/corr/HuangLW16a}
G.~{Huang}, Z.~{Liu}, L.~v.~d. {Maaten}, and K.~Q. {Weinberger}, ``Densely
  connected convolutional networks,'' in \emph{2017 IEEE Conference on Computer
  Vision and Pattern Recognition (CVPR)}, July 2017, pp. 2261--2269.

\bibitem{DBLP:journals/corr/abs-1805-02628}
\BIBentryALTinterwordspacing
M.~Juuti, S.~Szyller, A.~Dmitrenko, S.~Marchal, and N.~Asokan, ``{PRADA:}
  protecting against {DNN} model stealing attacks,'' \emph{CoRR}, vol.
  abs/1805.02628, 2018. [Online]. Available:
  \url{http://arxiv.org/abs/1805.02628}
\BIBentrySTDinterwordspacing

\bibitem{kesarwani2017model}
\BIBentryALTinterwordspacing
M.~Kesarwani, B.~Mukhoty, V.~Arya, and S.~Mehta, ``Model extraction warning in
  mlaas paradigm,'' in \emph{Proceedings of the 34th Annual Computer Security
  Applications Conference}, ser. ACSAC '18.\hskip 1em plus 0.5em minus
  0.4em\relax New York, NY, USA: ACM, 2018, pp. 371--380. [Online]. Available:
  \url{http://doi.acm.org/10.1145/3274694.3274740}
\BIBentrySTDinterwordspacing

\bibitem{Kocher:1996:TAI:646761.706156}
\BIBentryALTinterwordspacing
P.~C. Kocher, ``Timing attacks on implementations of diffie-hellman, rsa, dss,
  and other systems,'' in \emph{Proceedings of the 16th Annual International
  Cryptology Conference on Advances in Cryptology}, ser. CRYPTO '96.\hskip 1em
  plus 0.5em minus 0.4em\relax London, UK, UK: Springer-Verlag, 1996, pp.
  104--113. [Online]. Available:
  \url{http://dl.acm.org/citation.cfm?id=646761.706156}
\BIBentrySTDinterwordspacing

\bibitem{Kocher:1999:DPA:646764.703989}
\BIBentryALTinterwordspacing
P.~C. Kocher, J.~Jaffe, and B.~Jun, ``Differential power analysis,'' in
  \emph{Proceedings of the 19th Annual International Cryptology Conference on
  Advances in Cryptology}, ser. CRYPTO '99.\hskip 1em plus 0.5em minus
  0.4em\relax Berlin, Heidelberg: Springer-Verlag, 1999, pp. 388--397.
  [Online]. Available: \url{http://dl.acm.org/citation.cfm?id=646764.703989}
\BIBentrySTDinterwordspacing

\bibitem{cifar10}
\BIBentryALTinterwordspacing
A.~Krizhevsky, V.~Nair, and G.~Hinton, ``Cifar-10 (canadian institute for
  advanced research).'' [Online]. Available:
  \url{http://www.cs.toronto.edu/~kriz/cifar.html}
\BIBentrySTDinterwordspacing

\bibitem{NIPS2012_4824}
A.~Krizhevsky, I.~Sutskever, and G.~E. Hinton, ``Imagenet classification with
  deep convolutional neural networks,'' in \emph{Advances in Neural Information
  Processing Systems 25}, F.~Pereira, C.~J.~C. Burges, L.~Bottou, and K.~Q.
  Weinberger, Eds.\hskip 1em plus 0.5em minus 0.4em\relax Curran Associates,
  Inc., 2012, pp. 1097--1105.

\bibitem{milli2018model}
\BIBentryALTinterwordspacing
S.~Milli, L.~Schmidt, A.~D. Dragan, and M.~Hardt, ``Model reconstruction from
  model explanations,'' in \emph{Proceedings of the Conference on Fairness,
  Accountability, and Transparency}, ser. FAT* '19.\hskip 1em plus 0.5em minus
  0.4em\relax New York, NY, USA: ACM, 2019, pp. 1--9. [Online]. Available:
  \url{http://doi.acm.org/10.1145/3287560.3287562}
\BIBentrySTDinterwordspacing

\bibitem{Naghibijouybari:2018:RIG:3243734.3243831}
\BIBentryALTinterwordspacing
H.~Naghibijouybari, A.~Neupane, Z.~Qian, and N.~Abu-Ghazaleh, ``Rendered
  insecure: Gpu side channel attacks are practical,'' in \emph{Proceedings of
  the 2018 ACM SIGSAC Conference on Computer and Communications Security}, ser.
  CCS '18.\hskip 1em plus 0.5em minus 0.4em\relax New York, NY, USA: ACM, 2018,
  pp. 2139--2153. [Online]. Available:
  \url{http://doi.acm.org/10.1145/3243734.3243831}
\BIBentrySTDinterwordspacing

\bibitem{oh2018towards}
S.~J. Oh, M.~Augustin, M.~Fritz, and B.~Schiele, ``Towards reverse-engineering
  black-box neural networks,'' 2018.

\bibitem{orekondy19knockoff}
T.~Orekondy, B.~Schiele, and M.~Fritz, ``Knockoff nets: Stealing functionality
  of black-box models,'' in \emph{CVPR}, 2019.

\bibitem{Papernot:2017:PBA:3052973.3053009}
\BIBentryALTinterwordspacing
N.~Papernot, P.~McDaniel, I.~Goodfellow, S.~Jha, Z.~B. Celik, and A.~Swami,
  ``Practical black-box attacks against machine learning,'' in
  \emph{Proceedings of the 2017 ACM on Asia Conference on Computer and
  Communications Security}, ser. ASIA CCS '17.\hskip 1em plus 0.5em minus
  0.4em\relax New York, NY, USA: ACM, 2017, pp. 506--519. [Online]. Available:
  \url{http://doi.acm.org/10.1145/3052973.3053009}
\BIBentrySTDinterwordspacing

\bibitem{DBLP:journals/corr/abs-1904-01067}
\BIBentryALTinterwordspacing
A.~Salem, A.~Bhattacharyya, M.~Backes, M.~Fritz, and Y.~Zhang, ``Updates-leak:
  Data set inference and reconstruction attacks in online learning,''
  \emph{CoRR}, vol. abs/1904.01067, 2019. [Online]. Available:
  \url{http://arxiv.org/abs/1904.01067}
\BIBentrySTDinterwordspacing

\bibitem{salem2018ml}
\BIBentryALTinterwordspacing
A.~Salem, Y.~Zhang, M.~Humbert, P.~Berrang, M.~Fritz, and M.~Backes,
  ``Ml-leaks: Model and data independent membership inference attacks and
  defenses on machine learning models,'' in \emph{26th Annual Network and
  Distributed System Security Symposium (NDSS 2019)}, February 2019. [Online].
  Available: \url{https://publications.cispa.saarland/2754/}
\BIBentrySTDinterwordspacing

\bibitem{shokri2017membership}
R.~Shokri, M.~Stronati, C.~Song, and V.~Shmatikov, ``Membership inference
  attacks against machine learning models,'' in \emph{Security and Privacy
  (SP), 2017 IEEE Symposium on}, 2017.

\bibitem{inproceedings}
J.~Silva, R.~Berriel, C.~Badue, A.~De~Souza, and T.~Oliveira-Santos, ``Copycat
  cnn: Stealing knowledge by persuading confession with random non-labeled
  data,'' 05 2018.

\bibitem{DBLP:journals/corr/SimonyanZ14a}
\BIBentryALTinterwordspacing
K.~Simonyan and A.~Zisserman, ``Very deep convolutional networks for
  large-scale image recognition,'' \emph{CoRR}, vol. abs/1409.1556, 2014.
  [Online]. Available: \url{http://arxiv.org/abs/1409.1556}
\BIBentrySTDinterwordspacing

\bibitem{DBLP:journals/corr/SpringenbergDBR14}
\BIBentryALTinterwordspacing
J.~Springenberg, A.~Dosovitskiy, T.~Brox, and M.~Riedmiller, ``Striving for
  simplicity: The all convolutional net,'' in \emph{ICLR (workshop track)},
  2015. [Online]. Available:
  \url{http://lmb.informatik.uni-freiburg.de/Publications/2015/DB15a}
\BIBentrySTDinterwordspacing

\bibitem{Stefanov:2013:POE:2508859.2516660}
\BIBentryALTinterwordspacing
E.~Stefanov, M.~van Dijk, E.~Shi, C.~Fletcher, L.~Ren, X.~Yu, and S.~Devadas,
  ``Path oram: An extremely simple oblivious ram protocol,'' in
  \emph{Proceedings of the 2013 ACM SIGSAC Conference on Computer \&\#38;
  Communications Security}, ser. CCS '13.\hskip 1em plus 0.5em minus
  0.4em\relax New York, NY, USA: ACM, 2013, pp. 299--310. [Online]. Available:
  \url{http://doi.acm.org/10.1145/2508859.2516660}
\BIBentrySTDinterwordspacing

\bibitem{DBLP:journals/corr/SzeCYE17}
V.~{Sze}, Y.~{Chen}, T.~{Yang}, and J.~S. {Emer}, ``Efficient processing of
  deep neural networks: A tutorial and survey,'' \emph{Proceedings of the
  IEEE}, vol. 105, no.~12, pp. 2295--2329, Dec 2017.

\bibitem{tramer2016stealing}
F.~Tram{\`e}r, F.~Zhang, A.~Juels, M.~K. Reiter, and T.~Ristenpart, ``Stealing
  machine learning models via prediction apis,'' in \emph{USENIX Security},
  2016.

\bibitem{wang2018stealing}
B.~{Wang} and N.~Z. {Gong}, ``Stealing hyperparameters in machine learning,''
  in \emph{2018 IEEE Symposium on Security and Privacy (SP)}, May 2018, pp.
  36--52.

\bibitem{wei2018know}
\BIBentryALTinterwordspacing
L.~Wei, B.~Luo, Y.~Li, Y.~Liu, and Q.~Xu, ``I know what you see: Power
  side-channel attack on convolutional neural network accelerators,'' in
  \emph{Proceedings of the 34th Annual Computer Security Applications
  Conference}, ser. ACSAC '18.\hskip 1em plus 0.5em minus 0.4em\relax New York,
  NY, USA: ACM, 2018, pp. 393--406. [Online]. Available:
  \url{http://doi.acm.org/10.1145/3274694.3274696}
\BIBentrySTDinterwordspacing

\bibitem{Williams1992}
\BIBentryALTinterwordspacing
R.~J. Williams, ``Simple statistical gradient-following algorithms for
  connectionist reinforcement learning,'' \emph{Machine Learning}, vol.~8,
  no.~3, pp. 229--256, May 1992. [Online]. Available:
  \url{https://doi.org/10.1007/BF00992696}
\BIBentrySTDinterwordspacing

\bibitem{Xu2016AutomaticallyEC}
W.~Xu, Y.~Qi, and D.~Evans, ``Automatically evading classifiers: A case study
  on pdf malware classifiers,'' in \emph{NDSS}, 2016.

\bibitem{yan2018cache}
M.~Yan, C.~Fletcher, and J.~Torrellas, ``Cache telepathy: Leveraging shared
  resource attacks to learn dnn architectures,'' \emph{arXiv preprint
  arXiv:1808.04761}, 2018.

\bibitem{DBLP:journals/corr/ZophL16}
\BIBentryALTinterwordspacing
B.~Zoph and Q.~V. Le, ``Neural architecture search with reinforcement
  learning,'' \emph{CoRR}, vol. abs/1611.01578, 2016. [Online]. Available:
  \url{http://arxiv.org/abs/1611.01578}
\BIBentrySTDinterwordspacing

\end{thebibliography}
}

\end{document}